\documentclass[conference, a4paper]{IEEEtran}

\IEEEoverridecommandlockouts 
\usepackage{booktabs}
\usepackage{graphicx}
\usepackage{multirow, makecell}

\usepackage{amsmath,amssymb,latexsym}
\usepackage{rotating}
\usepackage{lettrine}
\usepackage{textcomp}
\usepackage{units}
\usepackage[table]{xcolor}
\usepackage{alltt}
\usepackage{hyperref}
\usepackage{lipsum}%
\usepackage[pscoord]{eso-pic}
\newcommand{\placetextbox}[3]{%
  \setbox0=\hbox{#3}
  \AddToShipoutPictureFG*{
    \put(\LenToUnit{#1\paperwidth},\LenToUnit{#2\paperheight}){\vtop{{\null}\makebox[0pt][c]{#3}}}%
  }%
}%

\usepackage{tikz}

\usepackage{xcolor}

\title{Wireless Sensor Networks based on TSCH/TDMA with Power Consumption and Latency Constraints
\thanks{This work was partially supported by the European Union under the Italian National Recovery and Resilience Plan (NRRP) of NextGenerationEU, partnership on ``Telecommunications of the Future'' (PE00000001 - program ``RESTART'').}
}

\author{
    \IEEEauthorblockN{
    Stefano Scanzio\IEEEauthorrefmark{1},
    Gabriele Formis\IEEEauthorrefmark{1}\IEEEauthorrefmark{2},
    Tullio Facchinetti\IEEEauthorrefmark{3},
    Giacomo Paolini\IEEEauthorrefmark{4},
    Gianluca Cena\IEEEauthorrefmark{1}
    }
    
    \IEEEauthorblockA{\IEEEauthorrefmark{1}National Research Council of Italy (CNR--IEIIT), Italy. \IEEEauthorrefmark{2}Politecnico di Torino, Italy.}    
    \IEEEauthorblockA{\IEEEauthorrefmark{3}University of Pavia, Italy. \IEEEauthorrefmark{4}University of Bologna, Italy.}
    Emails: stefano.scanzio@cnr.it, gabriele.formis@polito.it, tullio.facchinetti@unipv.it,\\ giacomo.paolini4@unibo.it, gianluca.cena@cnr.it
    }

\begin{document}
\placetextbox{0.5}{1}{This is the author's version of an article that has been published.}
\placetextbox{0.5}{0.985}{Changes were made to this version by the publisher prior to publication.}
\placetextbox{0.5}{0.97}{The final version of record is available at \href{https://doi.org/10.1109/ETFA61755.2024.10710912}{https://doi.org/10.1109/ETFA61755.2024.10710912}}%
\placetextbox{0.5}{0.05}{Copyright (c) 2024 IEEE. Personal use is permitted.}
\placetextbox{0.5}{0.035}{For any other purposes, permission must be obtained from the IEEE by emailing pubs-permissions@ieee.org.}%

\maketitle
\thispagestyle{empty}
\pagestyle{empty}

\begin{abstract}
One of the main goals of wireless sensor networks is to permit the involved nodes to communicate with low energy budgets, as they are typically battery-powered. 
When such networks are employed in industrial scenarios, constraints about latency may have a significant role, too. 
The TSCH mechanism, and more in general TDMA schemes, rely on traffic scheduling, and consequently they can feature low power consumption and more predictable latency. 
Some recent proposals like \mbox{PRIL-M} enable further consistent energy savings, but unfortunately they cause at the same time a dramatic increase in latency.

This work presents an extension of PRIL-M, we named \mbox{PRIL-ML}, that achieves a significantly shorter latency in exchange for a slight increase in power consumption.
Its operating principles are first illustrated, then some approximate equations are provided for assessing analytically the improvements it achieves, starting from simulation results obtained for both standard TSCH and the original PRIL-M technique.
\end{abstract}


\section{Introduction}
\label{sec:introduction}
Wireless sensor networks (WSNs) are gaining ground in the industrial network landscape, which is characterized by a high degree of heterogeneity in terms of the employed communication protocols and the related characteristics \cite{SCANZIO2021103388}.
One of the
main objectives of WSNs is to achieve extremely low power consumption. 
Nonetheless, 
other objectives may exist as well. 
For example,
the ability to ensure predictable and bounded latency (although orders of magnitude larger than typical industrial protocols) 
may sometimes be
of non-negligible importance. 
The time slotted channel hopping (TSCH) \cite{8823863} protocol defined in the IEEE 802.15.4 standard \cite{IEEE-802.15.4-2020} relies on traffic scheduling \cite{10208136} and provides improvements concerning both power consumption and determinism \cite{9187609, CENA2020102199}. 
The basic TSCH access technique splits time into fixed-length slots and resembles time division multiple access (TDMA). 
Additionally, channel hopping makes TSCH more robust against narrowband disturbance \cite{8534544,10158374}.

Several techniques have been developed to enhance energy efficiency in TSCH-based WSNs. 
The approach in \cite{9730918} involves non-orthogonal multiple access (NOMA) use, which improves spectral efficiency, reduces latency, and increases energy efficiency by allowing sensor nodes to transmit data to their cluster heads using NOMA transmissions. 
Another technique is the 6TiSCH Low Latency Autonomous Scheduling \cite{9815595}, which reduces latency and radio duty cycle, thus enhancing energy efficiency and reliability in Industrial Internet of Things (IIoT) applications. 
Additionally, the MSU-TSCH algorithm \cite{9925697} supports mobility and adapts to topology changes, significantly reducing energy consumption and end-to-end latency. 
The DualBlock scheme \cite{9810842} introduces an adaptive intra-slot CSMA/CA mechanism, reducing collisions and energy consumption within congested slots.

Besides above techniques, Proactive Reduction of Idle Listening (PRIL) has been also proposed recently.
Its initial version (PRIL-F) \cite{9903301,10.1007/978-3-030-61746-2_11} permits to lower energy consumption on the first-hop of every path.
Subsequently, it has been extended to properly manage multi-hop networks as well (PRIL-M) \cite{2024-IoTJ}.
In practice, PRIL-F decreases power consumption of nodes located one hop away from the source node, 
whereas \mbox{PRIL-M} deals with all the following hops to the destination.
From the point of view of energy reduction, \mbox{PRIL-M} performance is optimal.
Unfortunately, it leads to a dramatic increase of communication latency, which
makes it unsuitable in several specific industrial contexts.

In this paper, 
PRIL multi-hop with latency constraints (PRIL-ML) 
is proposed, which 
is significantly more responsive than
PRIL-M
by slightly increasing energy consumption.
Due to the similarity between TSCH and TDMA, all the ideas described here can be easily applied to the latter too.

The paper organization is the following: 
in Section~\ref{sec:PRILM}, the TSCH protocol and the former PRIL-M technique are described, 
while the new PRIL-ML technique is introduced in Section~\ref{sec:PRILM2}. 
Results about TSCH, PRIL-M, and some analytical estimates about PRIL-ML are reported in Section~\ref{sec:results}, before Section~\ref{sec:conc} that concludes the work.

\section{TSCH and PRIL-M}
\label{sec:PRILM}
\begin{figure}[t]
    \begin{center}
    \includegraphics[width=1.0\columnwidth]{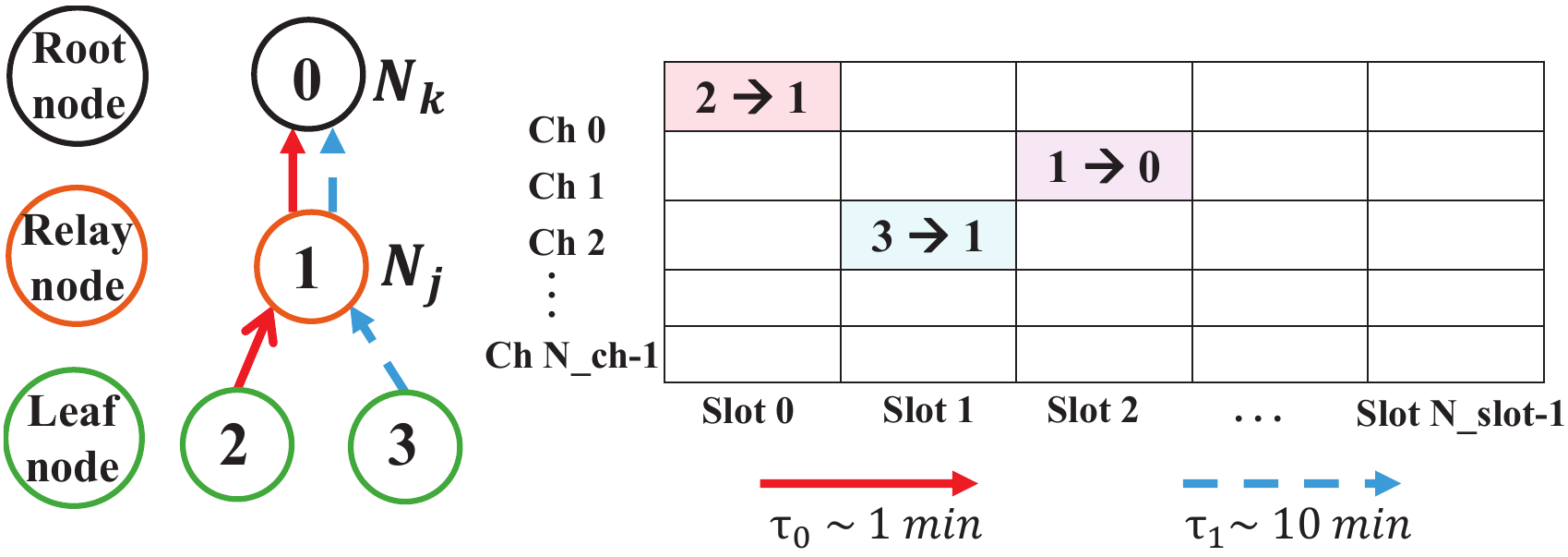}
    \end{center}

    \caption{Example of TSCH operation and slotframe matrix.}

    \label{fig:PRIL}
\end{figure}

As depicted in Fig.~\ref{fig:PRIL}, TSCH partitions the wireless transmission medium on both the time and frequency domains by means of a slotframe matrix.
Time partitioning is a sort of TDMA, where time is split into a number of fixed-size slots (often there are $101$, whose duration is typically either $10$ or $\unit[20]{ms}$). 
Frequency division enables instead concurrent transmissions (on up to $16$ channels in the $\unit[2.4]{GHz}$ band), which increase the overall network throughput.
Every cell in the matrix, identified by its slot and channel offsets, is allocated to one (or more) specific links.
Besides preventing collisions, traffic scheduling also leads to a consistent reduction of power consumption,
since nodes can stay in a deep-sleep state and wake up only in those slots where 
they must either transmit or receive a frame.
Additionally, channel hopping provides more stable links by decreasing 
variability
of the frame loss ratio
\cite{10158374}. 
In fact, the slotframe repeats over time (period is $\unit[2.02]{s}$ in the case of $101$ slots of width $\unit[20]{ms}$) and, at every repetition, the slot scheduled to the link is mapped to a different physical channel, providing both time and frequency diversity.

Fig.~\ref{fig:PRIL} depicts a multi-hop network of four nodes organized in two layers
(the tree routing topology is constructed, e.g., by the RPL protocol),
which is used as a running example
throughout this work. 
The global schedule is reported in the matrix on the right side of the figure.
Relevant portions of it are stored inside every node, which manage its possible states. 
For instance, $\mathrm{N}_1$ switches to the \textit{receive} state in slots $0$ and $1$, and in the \textit{transmission} state in slot $2$ (only if its queue contains at least one frame addressed to node $\mathrm{N}_0$).
As the figure shows, two periodic flows are defined, $\tau_0$ and $\tau_1$, which originate from leaf nodes and are directed to the root.
The nodes traversed by the packets of $\tau_0$ are $\mathrm{N}_2$, $\mathrm{N}_1$, and $\mathrm{N}_0$, 
while for $\tau_1$ they are $\mathrm{N}_3$, $\mathrm{N}_1$, and $\mathrm{N}_0$.
Due to tolerances and random drifts of the nodes' quartz oscillators, the generation processes of flows 
(defined in terms of their nominal rate and initial phase) are typically asynchronous.
Generation periods in our sample network were set to $T_{\tau_0} \simeq \unit[1]{min}$ and $T_{\tau_1} \simeq \unit[10]{min}$, 
but not necessarily they have to be multiples of each other. 
These aspects have been taken into account in this work.

The presence of flows with quite long (and different) periods is commonplace in WSNs. Since (at least) one transmission opportunity (i.e., cell) is configured in the slotframe matrix for any given link every $\unit[2.02]{s}$, 
while generations periods are
much larger
($\unit[1]{}$ and $\unit[10]{min}$), most of the reserved slots remain unused.
This makes receiving nodes waste a non-negligible amount of energy waiting for frames that will never arrive, a phenomenon known as \textit{idle listening}. 
The simulator was configured in such a way that the energy spent for sending a maximal-size frame ($\unit[127]{B}$), for receiving it, and wasted for idle listening is $E_{\mathrm{send}}=\unit[485.7]{\mu J}$, $E_{\mathrm{rec}}=\unit[651.0]{\mu J}$, and $E_{\mathrm{listen}}=\unit[303.3]{\mu J}$, respectively. 
These values are derived from the experimental results reported in \cite{2014-SJ-consumption}, related to an OpenMoteSTM device equipped with an Atmel AT86RF231 radiochip and an STM32F103RB 32-bit microcontroller.

\begin{figure}[t]
	\begin{center}
	\includegraphics[width=1.0\columnwidth]{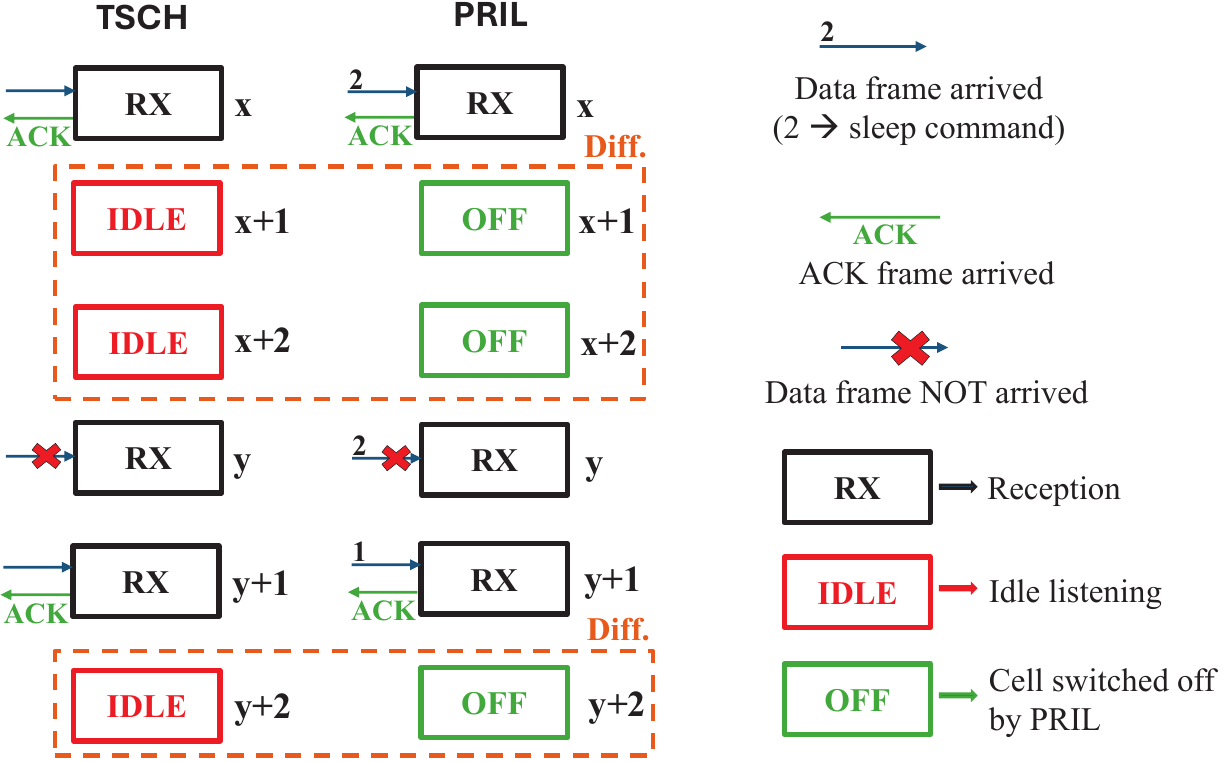}
	\end{center}
       \caption{Example of PRIL-F operation compared with default TSCH.}
	\label{fig:PRIL_operation}

\end{figure}

PRIL techniques rely on specific sleep commands, which are embedded in the frames and permit the sender to turn off the receiver for a given amount of time, in such a way to prevent (or just lower) the impact of idle listening. 
If only periodic traffic is considered, the receiving side of the link that corresponds to the first hop of the path followed by the packets of a given flow can be disabled precisely, without any impact on latency but with a significant decrease in energy consumption. 
This technique is termed PRIL-F and is described in \cite{9903301}.

The simple example in Fig.~\ref{fig:PRIL_operation} compares the behavior of PRIL-F and standard TSCH.
Rectangular shapes represent the instances of the slot allocated to the link in subsequent slotframes.
The packet generation period at the application level is 
set
equal to $3$ slotframes, i.e., $\unit[6.06]{s}$. 
In PRIL-F, the sleep command (with value $2$) added to the frame (at time $x$) disables the following two slots on the receiver (i.e., it sets their state to \texttt{OFF}), preventing idle listening completely. 
In the second exchange (at time $y$) the initial frame is assumed to be lost, and hence only one slot can be disabled by PRIL-F. 

PRIL-M has been conceived to complement PRIL-F (which only deals with the first hop),
and operates on the following hops (e.g., from $\mathrm{N}_1$ to $\mathrm{N}_0$ in Fig.~\ref{fig:PRIL}).
Since the involved links can be traversed by multiple flows with different periods, 
relay nodes set the duration of sleep commands based on the fastest flow they see, i.e., the one with the minimum period (in the example of Fig.~\ref{fig:PRIL}, $\tau_0$, whose period is $T_{\tau_0}=\unit[1]{min}$). 
This permits to prioritize this flow, which 
is expected to have more demanding timing requirements. 
On the one hand this leads to a significant reduction in energy consumption, but on the other it causes a substantial latency increase for the other flows. 
Detailed explanations about PRIL-M can be found in \cite{2024-IoTJ}.

\section{PRIL-ML}
\label{sec:PRILM2}
The basic idea behind the PRIL-ML technique, first presented in this paper, is to decrease the 
time for which the receiving side of a link is turned off by PRIL-M, to limit its impact on latency. 
A preliminary analysis carried out in this study highlights a noticeable decrease in latency, with a negligible increase in energy consumption. 
As for PRIL-M, PRIL-ML only acts starting from the second hop of a path.

Referring to Fig.~\ref{fig:PRIL} and focusing on the link \mbox{$j\rightarrow k$} between node $\mathrm{N}_j$ and its neighbor $\mathrm{N}_k$, PRIL-M permits $\mathrm{N}_j$ to disable those cells of $\mathrm{N}_k$ that are affected by idle listening. 
Among all the flows $\tau_i$ traversing it, node $\mathrm{N}_j$ considers those directed to $\mathrm{N}_k$ (a separate instance of PRIL-M is created for every outgoing link). 
Let $\mathcal{T}_{j\rightarrow k}=\{\tau_i|\tau_i\ \text{crosses}\ \mathrm{N}_j\ \text{and then}\ \mathrm{N}_k\}$ be the set of all the flows that traverse both $\mathrm{N}_j$ and $\mathrm{N}_k$,
in that order,
and $T_{\min}^{j\rightarrow k}=\min(T_{\tau_i} | \tau_i \in \mathcal{T}_{j\rightarrow k})$ the minimum period between flows belonging to $\mathcal{T}_{j\rightarrow k}$. 
The related (fastest) flow is identified with the symbol $\tau_*^{j\rightarrow k}$. 
The PRIL-M algorithm foresees that $\mathrm{N}_j$ periodically turns off 
part of the slots scheduled to link $j\rightarrow k$
in such a way to prevent idle listening on $\mathrm{N}_k$.
In particular, this procedure is triggered by the reception of packets from $\tau_*^{j\rightarrow k}$, which occurs with period $T_{\min}^{j\rightarrow k}$.

Since packet generation for the different flows is asynchronous, 
we can assume that a packet $p_i$ that belongs to $\tau_i$ ($p_i \in \tau_i$), with $\tau_i \ne \tau_*^{j\rightarrow k}$, 
arrives to node $\mathrm{N}_j$ (where it is enqueued for relay) at a time uniformly distributed in the interval 
$[0, T_{\min}^{j\rightarrow k}]$ between two subsequent packets belonging to $\tau_*^{j\rightarrow k}$. 
This means that, due to the way relay is performed by $\mathrm{N}_j$, 
the average latency of packet $p_i$ increases by $\Delta^{\mathrm{PRIL-M}}_{\mu} ={T_{\min}^{j\rightarrow k}}/{2}$,
while its worst-case latency grows by $\Delta^{\mathrm{PRIL-M}}_{\max}=T_{\min}^{j\rightarrow k}$ 
(this happens when $p_i$ arrives to $\mathrm{N}_j$ just after the packet belonging to $\tau_*^{j\rightarrow k}$ has been sent, which temporarily deactivates all the following cells that cause idle listening on receiver $\mathrm{N}_k$).
Above simplified formulas do not take into account the effect of retransmissions on latency and power consumption (which are instead modeled  by the simulator properly). 
Nevertheless, experimental data of Section~\ref{sec:results} show that they provide a good approximation of reality, and are adequate to evaluate the effect of the proposed  PRIL-ML technique.

The sleep command is included by $\mathrm{N}_j$ (encoded as a TSCH \textit{information element}) every time it relays the last packet in the $j\rightarrow k$ buffer.
Hence, upon the arrival of a packet from $\tau_*^{j\rightarrow k}$, all enqueued packets are always forwarded according to TSCH rules, before the link is switched off by PRIL.
The sleep duration in PRIL-M is derived from the value $T_{\min}^{j\rightarrow k}$, by subtracting the actual time taken to empty the buffer by sending all the queued frames.
Since transmission takes place in 
First-In First-Out order,
the packet from $\tau_*^{j\rightarrow k}$ can be preceded and followed 
in the queue
by packets from other flows.
The main idea behind PRIL-ML is to set the nominal sleep period $T_{\mathrm{act}}^{j\rightarrow k}$ of outgoing link $j\rightarrow k$ to a sub-multiple of $T_{\min}^{j\rightarrow k}$, that is, $T_{\mathrm{act}}^{j\rightarrow k} = \left \lceil {T_{\min}^{j\rightarrow k}}/{r} \right \rceil$. 
In theory, 
any duration $T_{\mathrm{act}}^{j\rightarrow k} \leq T_{\min}^{j\rightarrow k}$ could be chosen,
but we believe this does not add much from the applications' perspective. 
If, for instance, we set $r=4$, the link between $\mathrm{N}_j$ and $\mathrm{N}_k$ is reopened after a time equal to about $T_{\min}^{j\rightarrow k}/4$, 
allowing node $\mathrm{N}_j$ to flush any queued packets four times as fast. 

The increase in power consumption 
over PRIL-M
due to PRIL-ML adoption is equal, at worst, to $\Delta P = \frac{(r-1) \cdot E_{\mathrm{listen}}}{T_{\min}^{j\rightarrow k}}$, where $E_{\mathrm{listen}}$ is the energy wasted by the receiver when idle listening takes place in one slot. 
$\Delta P$ represents an upper bound because packets queued in $\mathrm{N}_j$ would be sent anyway.
Regarding latency, the increase in the relay time of $\mathrm{N}_j$ 
over TSCH
due to \mbox{PRIL-ML} adoption is equal, on average, to 
$\Delta^{\mathrm{PRIL-ML}}_{\mu} =  {T_{\mathrm{act}}^{j\rightarrow k}}/{2} $, 
while in the worst case it grows by $\Delta^{\mathrm{PRIL-ML}}_{\mathrm{max}} = {T_{\mathrm{act}}^{j\rightarrow k}}$. 
In fact, every time the outgoing link $j\rightarrow k$ is re-opened, packets enqueued in  $\mathrm{N}_j$ can be sent without further delays.
This means that latency is $r$ times shorter than PRIL-M.

Regarding the implementation of PRIL-ML, when a frame delivering an extended sleep message (bearing both $T_{\mathrm{act}}$ and $r$) arrives on the receiving side of link $j\rightarrow k$, it triggers a sequence of actions in node $\mathrm{N}_k$.
In particular, for $r$ times the receiver is put to sleep for $T_{\mathrm{act}}^{j\rightarrow k}$ (the final sleeping time may be slightly shorter),  awakening it every time for (at least) one slot, 
as if flow $\tau_*^{j\rightarrow k}$ were $r$ time faster than it actually is.
PRIL-ML complexity is reasonably low \cite{2024-IoTJ}, which enables inclusion in inexpensive motes.

\section{Results}
\label{sec:results}
The TSCHmodeler discrete-event simulator was used to simulate the network in Fig.~\ref{fig:PRIL}
with both standard TSCH and PRIL-M
(it has not been extended to PRIL-ML yet).
Every simulation has a duration of $\unit[10]{years}$, and the slotframe matrix was configured so that the first three slots are allocated to transmissions from $\mathrm{N}_2$ to $\mathrm{N}_1$, from $\mathrm{N}_3$ to $\mathrm{N}_1$, and from $\mathrm{N}_1$ to $\mathrm{N}_0$, respectively. 
An experimental campaign was preliminary carried out (not reported here for space reasons)
that shows 
that the results provided by the simulator for a TSCH network 
are very similar to those obtained on real OpenMote~B 
devices
running OpenWSN (6TiSCH).

\begin{table}[t]
\caption{Energy consumption for the different techniques and nodes}
\vspace{-0.2cm}
\label{tab:power}
\centering
\renewcommand{\arraystretch}{1.0}
\begin{tabular}{clrrrrr}
\toprule
Technique & Power ($\mu W$) & $\mathrm{N}_{0}$ & $\mathrm{N}_{1}$ & $\mathrm{N}_{2}$ & $\mathrm{N}_{3}$ & $\mathrm{All}$ \\
\midrule
\multirowcell{4}{\rotatebox[origin=c]{90}{TSCH}}
& $P_{\mathrm{send}}$ & 0.0 & 11.1 & 10.1 & 1.0 & 22.2 \\
& $P_{\mathrm{rec}}$ & 14.8 & 14.9 & 0.0 & 0.0 & 29.7 \\
& $P_{\mathrm{listen}}$ & 143.2 & 293.4 & 0.0 & 0.0 & 436.6 \\
& $P$ & 158.0 & 319.4 & 10.1 & 1.0 & 488.5 \\
\midrule
\multirowcell{4}{\rotatebox[origin=c]{90}{PRIL-M}}
& $P_{\mathrm{send}}$ & 0.0 & 19.9 & 18.9 & 1.9 & 40.7 \\
& $P_{\mathrm{rec}}$ & 13.8 & 13.7 & 0.0 & 0.0 & 27.5 \\
& $P_{\mathrm{listen}}$ & 0.4 & 0.0 & 0.0 & 0.0 & 0.4 \\
& $P$ & 14.2 & 33.6 & 18.9 & 1.9 & 68.6 \\
\bottomrule
\end{tabular}
\vspace{-0.3cm}
\end{table}

\begin{table}[t]
\caption{End-to-end latency for the different techniques and flows}
\vspace{-0.2cm}
\label{tab:latency}
\centering
\tabcolsep=0.11cm	
\renewcommand{\arraystretch}{1.0}
\begin{tabular}{clrrrrrrr}
\toprule
Tech. & Latency ($s$) & $\mu_d$ & $\sigma_d$ & $d_{\min}$ & $d_{p99}$ & $d_{p99.9}$ & $d_{p99.99}$ & $d_{\mathrm{max}}$ \\
\midrule
\multirowcell{3}{\rotatebox[origin=c]{90}{TSCH}}
& $\tau_0$ & 1.644 & 1.300 & 0.060 & 5.960 & 8.360 & 11.12 & 18.12 \\
& $\tau_1$ & 1.731 & 1.413 & 0.040 & 6.440 & 9.500 & 11.90 & 17.96 \\
& $\tau_0 + \tau_1$ & 1.652 & 1.310 & 0.040 & 5.980 & 8.560 & 11.28 & 18.12 \\
\midrule
\multirowcell{3}{\rotatebox[origin=c]{90}{PRIL-M}}
& $\tau_0$ & 2.658 & 1.587 & 0.060 & 7.940 & 10.64 & 13.72 & 22.00 \\
& \textbf{$\tau_1$} & \textbf{30.58} & \textbf{17.20} & \textbf{0.040} & \textbf{60.38} & \textbf{63.20} & \textbf{65.94} & \textbf{69.86} \\
& $\tau_0 + \tau_1$ & 5.196 & 9.676 & 0.040 & 53.86 & 60.30 & 63.04 & 69.86 \\
\bottomrule
\end{tabular}
\vspace{-0.3cm}
\end{table}

In the case of periodic traffic with a known generation period, the effect of PRIL-M on energy consumption is extremely positive, as shown by the results reported in Table~\ref{tab:power}. 
As can be seen, the idle listening component of energy consumption is practically zero for all nodes in the network.
Unfortunately, PRIL-M impacts negatively on end-to-end latency, as shown in Table~\ref{tab:latency}, which reports the main statistics 
about flows $\tau_0$ and $\tau_1$, namely
average value ($\mu_d$), standard deviation ($\sigma_d$), minimum ($d_{\min}$), 
99-percentile ($d_{p99})$, 99.9-percentile ($d_{p99.9}$), 99.99-percentile ($d_{p99.99})$, and maximum ($d_{\mathrm{max}}$), for both standard TSCH and PRIL-M.

Regarding PRIL-M, latency increase for the fastest flow $\tau_0$ 
over TSCH
is rather small compared to $\tau_1$. 
This
can be explained by
looking at the example of Fig.~\ref{fig:PRIL}.
Any retransmissions of a packet
belonging to $\tau_0$
on the link $2\rightarrow 1$ 
cause a delay on its arrival on $\mathrm{N}_1$. 
If the queue of $\mathrm{N}_1$ is empty, the packet will be forwarded with a sleep command equal to $T_{\min}^{1\rightarrow 0} = T_{\tau_0}  \simeq \unit[1]{min}$, 
otherwise a more complex algorithm is executed as described in \cite{2024-IoTJ}. 
If the next packet belonging to 
$\tau_0$, which is 
generated by
$\mathrm{N}_2$ after a time $T_{\tau_0}$, arrives to node $\mathrm{N}_1$ without retransmissions, 
it finds the link $1\rightarrow 0$ in the \texttt{OFF} state and suffers from an additional delay that increases 
latency beyond what the packet would experience without PRIL-M.

The main problem concerns, however, the 
steep
latency increase for flow $\tau_1$, whose statistics 
in Table~\ref{tab:latency}
are set in boldface.
The simulated average latency for \mbox{PRIL-M} is $\mu^{\mathrm{PRIL-M}}_{d} = \unit[30.58]{s}$, 
which resembles the value obtained
with our approximate model 
by summing 
$\Delta^{\mathrm{PRIL-M}}_{\mu} = T_{\min}^{1\rightarrow 0}/2$ to the average latency of TSCH, 
that is, 
$\hat\mu^{\mathrm{PRIL-M}}_{d} = \mu^{\mathrm{TSCH}}_{d} + \Delta^{\mathrm{PRIL-M}}_{\mu} = 1.731 + 30 = \unit[31.731]{s}$. 
A similar reasoning can be done to 
estimate the
maximum latency as
$\hat d^{\mathrm{PRIL-M}}_{\mathrm{max}} = d^{\mathrm{TSCH}}_{\mathrm{max}} + \Delta^{\mathrm{PRIL-M}}_{\mathrm{max}} = 17.96 + 60 = \unit[77.96]{s}$.
In this case, the theoretical value differs in a more pronounced way from the simulated one
$d^{\mathrm{PRIL-M}}_{\mathrm{max}}=\unit[69.86]{s}$,
as the maximum is more sensitive to the actual occurrence of rare events
(for example, having a high number of retransmissions for the same packet on both links $3\rightarrow 1$ and $1\rightarrow 0$).

At present, we are not able to simulate \mbox{PRIL-ML} behavior yet.
Nevertheless,
we can apply our simple theoretical model to find satisfactory estimates.
By setting $r=4$, we obtain 
$\Delta^{\mathrm{PRIL-ML}}_{\mu}=\unit[7.5]{s}$ and 
$\Delta^{\mathrm{PRIL-ML}}_{\mathrm{max}}=\unit[15]{s}$.
Consequently, approximated values for the average and worst-case latency 
can be evaluated as
$\hat\mu^{\mathrm{PRIL-ML}}_{d} = \unit[9.231]{s} $ and 
$\hat d^{\mathrm{PRIL-ML}}_{\mathrm{max}} = \unit[32.96]{s}$, respectively.
The price to be paid for latency decrease is the rise of power consumption. 
Luckily, this increase is fairly limited compared to the total power consumption of TSCH ($P^{\mathrm{TSCH}}=\unit[488.5]{\mu W}$), 
and can be evaluated as 
$\Delta P = {(4-1)\cdot \unit[303.3]{\mu J}}/{\unit[60]{s}} = \unit[15.2]{\mu W}$. 

This implies that 
the overall power consumption of the entire network 
when \mbox{PRIL-ML} is employed is approximately
$\hat P^{\mathrm{PRIL-ML}} = P^{\mathrm{PRIL-M}}+\Delta P = 68.6+15.2 = \unit[83.8]{\mu W}$. 
Such increase only affects the idle listening component of node $\mathrm{N}_0$. 
By using our theoretical model for PRIL-ML, 
this specific contribution
grows from $P^{\mathrm{PRIL-M}}_{\mathrm{listen}}=\unit[0.4]{\mu W}$ up to 
$\hat P^{\mathrm{PRIL-ML}}_{\mathrm{listen}}=0.4+15.2=\unit[15.6]{\mu W}$, which remains quite small compared to TSCH, where
$P^{\mathrm{TSCH}}_{\mathrm{listen}}=\unit[143.2]{\mu W}$.

\section{conclusion}
\label{sec:conc}
A new improvement for TSCH networks, which was named PRIL-ML, has been proposed in this paper to address the main drawback of the former PRIL-M technique, that is, a significant increase in communication latency. 
Although \mbox{PRIL-M} succeeds in minimizing energy consumption as much as possible, its considerably high delays may prevent its adoption in time-sensitive applications.

PRIL-ML permits to adjust the trade-off between energy consumption and end-to-end latency,
and consequently it makes it possible to customize the network parameters based on the application requirements.
For instance, 
the average latency 
can be shrunk
to 
less than
one-third of PRIL-M with an increase in the total power consumption of just $22\%$.
As future work we plan to extend the simulator to tackle \mbox{PRIL-ML}, 
so that it can be studied in detail.

\bibliographystyle{IEEEtran}
\bibliography{bibliography}

\end{document}